\documentclass[preprint,showpacs,preprintnumbers,amsmath,amssymb]{revtex4}

\usepackage{graphicx}
\usepackage{dcolumn}
\usepackage{bm}

\begin{document}

\preprint{}

\title{Space- and time-like electromagnetic pion form factors in light-cone pQCD}

\author{Jiunn-Wei Chen}
 \email{jwc@phys.ntu.edu.tw}
 \affiliation{
Department of Physics and Center for Theoretical Sciences,
National Taiwan University, Taipei 10617, Taiwan
}%
\author{Hiroaki Kohyama}%
 \email{kohyama@phys.sinica.edu.tw}
 \affiliation{%
Institute of Physics, Academia Sinica, Taipei, Taiwan 115, Republic of China
}
\affiliation{%
Physics Division, National Center for Theoretical Sciences,
Hsinchu, Taiwan 300, Republic of China
}%
\author{Kazuaki Ohnishi}%
 \email{kohnishi@phys.ntu.edu.tw}
 \affiliation{%
Department of Physics and Center for Theoretical Sciences,
National Taiwan University, Taipei 10617, Taiwan
}%
\author{Udit Raha}%
 \email{udit@phys.ntu.edu.tw}
 \affiliation{%
Department of Physics and Center for Theoretical Sciences,
National Taiwan University, Taipei 10617, Taiwan
}%
\author{Yue-Long Shen}%
 \email{shenyl@phys.sinica.edu.tw}
 \affiliation{%
Institute of Physics, Academia Sinica, Taipei, Taiwan 115, Republic of China
}%



\begin{abstract}
\noindent We present a combined analysis of the space- and time-like
electromagnetic pion form factors in light-cone perturbative QCD with transverse
momentum dependence and Sudakov suppression. Including the non-perturbative
``soft' QCD and power suppressed twist-3 corrections to the standard twist-2 
perturbative QCD result, the experimental pion data available at moderate
energies/momentum transfers can be explained reasonably well. This may help 
towards resolving the bulk of the existing discrepancy between the
space- and time-like experimental data. 
\end{abstract}


\maketitle

\noindent {\it Introduction.} The electromagnetic (e.m.) form factor of hadrons are
important physical observables that play a key role in
understanding the transition from the perturbative to the
non-perturbative behavior in particle physics. The space- and time-like
e.m. pion form factors $F_\pi$ and $G_\pi$, respectively,
are specified through the following matrix elements:
\begin{eqnarray}
e(P '+P)_\mu\, F_\pi(Q^2)\!\!&=&\!\!\left<\pi^\pm(P ')\left|J^{\rm em}_\mu(0)\right|\pi^\pm(P)\right>\,,\nonumber\\
e(P '-P)_\mu\, G_\pi(Q^2)\!\!&=&\!\!\left<\pi^+(P ')\pi^-(P)\left|J^{\rm em}_\mu(0)\right|0\right>\,,
\end{eqnarray}
where $J^{\rm em}_\mu$ is the e.m. current, and
$P\!=\!(Q/\sqrt{2},0,{\bf 0}_{T})$ and $P'\!=\!(0,Q/\sqrt{2},{\bf
  0}_{T})$ are, respectively, the initial and final state external light-cone pion
4-momenta in the Breit-frame. For the space-like momentum transfers, $q^2\!=\!(P'-P)^2\!=\!-Q^2\leq 0$,
whereas for the time-like momentum transfers $q^2\!=\!(P'+P)^2\!=\!Q^2\geq 0$.

Theoretical predictions based on standard \textquotedblleft
asymptotic\textquotedblright\, QCD rely on
collinear factorization \cite{CollFactor} that lead to
the celebrated quark counting rule, $\{F,G\}_\pi(Q^2)\sim 1/Q^2$ \cite{BF}.
Naively, one may then conclude that at high enough energies/momentum
transfers the space- and time-like form factors are essentially
of the same magnitude. However, at the present experimentally
accessible energies the reported pion form factor results
differ significantly, with the time-like results \cite{ee-hh1,ee-hh2,CLEO}
being up to a factor of four more than the space-like results \cite{pion_exp}.
Efforts to explain the above difference with conventional Vector Meson
Dominance (VMD) \cite{pion_exp} and perturbative QCD (pQCD) lead to the general
conclusion that the time-like and the space-like data are inconsistent
with each other. The purpose of this letter, is to show a possible ``clean''
scenario where the above difference could be reconciled with the standard
treatment of the parton transverse momentum dependence (TMD), the sub-leading twist-3
contributions and the so-called \textquotedblleft soft\textquotedblright\, QCD corrections.
However, in dealing with the parton picture in pQCD, one should naturally
be aware of the fact that in reality there are additional difficulties
with hadronization and other final state interactions, and resonances. 
Our approach assumes that these effects for $Q^2\gg
\Lambda^2_{\rm QCD}$ are comparatively small (as data suggest)
and would not come in conflict with our predictions that account
for the largeness of the existing discrepancy between the space-and
time-like pion form factor data.

The charged pion form factor can be written as \cite{Bakulev}
$\{F,G\}_\pi(Q^2)=\{F,G\}^{\rm soft}_\pi(Q^2)+\{F,G\}^{\rm hard}_\pi(Q^2)$.
The factorizable hard part $\{F,G\}^{\rm hard}_\pi(Q^2)$ is calculated
using light-cone pQCD with explicit TMD of the constituent valence partons; whereas, the
non-factorizable soft part $\{F,G\}^{\rm soft}_\pi(Q^2)$ is modeled using QCD
sum rules (QCDSR) via local quark-hadron duality.

Parametrically, both the soft and higher twist contributions to the form factor are
expected to be small at large momentum transfers compared to the leading hard
(twist-2) contributions due to the relative $1/Q^{2n}$ suppression. Despite
this, their contributions turn out to be unnaturally large at moderate
range of energies. In this paper, for the first time, we show that the twist-3
corrections to the time-like pion form factor are very large and essentially account
for the bulk of the observed discrepancy between theory and experimental data. Note that
the first attempt to explain both the space- and time-like data in the context
of pQCD includes only the twist-2 effects \cite{Pire}. However, the present
consensus is that the twist-2 effects are much too small to explain
the form factor data \cite{Bakulev,KLS,Wei,Huang,Raha}. Furthermore, one must
use appropriate Sudakov factors \cite{St,LiSt,Stefanis,Sanda,Li} to suppress
the kinematic enhancements that may invalidate factorization. The advantage of 
such a modified `$k_T$'-factorization \cite{St,LiSt,kT}
approach is the elimination of large logarithms in the hard kernel through the TMD
of the valence partons. This extends the range of applicability of pQCD down
to very moderate range of energies and has been widely applied to inclusive and
exclusive processes, and especially, to exclusive $B$-meson decays
\cite{KLS,Sanda,Li,B-decay}.

\begin{figure}[t]
    \hspace{0cm}
      \resizebox{10cm}{!}{\includegraphics{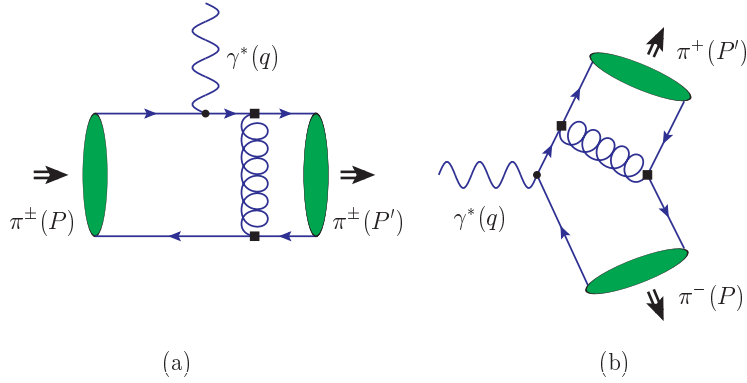}}
    \caption{LO representative matrix elements contributing to
            (a) the space-like $F_\pi(Q^2)$,  and (b) the time-like $G_\pi(Q^2)$
            pion form factors. The blobs represent the pion wavefunctions
            $\tilde{{\mathcal P}}_{\pi}$.}
    \label{fig:feynman}
\end{figure}

\noindent {\it Factorized pQCD.\,} We now present the essentials of our calculations.
The dominant contributions come only from the leading order (LO) Fock state,
i.e., a $q\bar{q}$ valence quark configuration with one hard gluon exchange in
the scattering kernel sandwiched between 2-particle wavefunctions/distribution
amplitudes (DAs). One of four diagrams contributing to each of the Born
amplitudes $\pi\gamma^*\rightarrow\pi$ and $\gamma^*\rightarrow\pi^+\pi^-$ is
shown in Fig.~\ref{fig:feynman}. The other diagrams correspond to allowing
the gluon to interact on the other side of the photon vertex and allowing the
photon also to couple to the other valence quark. The higher Fock state
contributions are neglected being suppressed by higher powers of $1/Q^2$.
Since, we are only concerned with the leading Fock states in the scattering
kernel of the pion, we must consider only the 2-particle pion DAs for our
analysis neglecting the multi-particle components.
Nevertheless, one can show that for the 2-particle twist-3 DAs are
not independent of the 3-particle twist-3 DA, being related by QCD equations
of motions. To next-to-leading order in conformal twist there is just one
2-particle twist-2 collinear DA $\phi_{2;\pi}(x,\mu)$ with an axial-vector
structure, and two 2-particle twist-3 collinear DAs, one with a pseudo-scalar
structure $\phi^p_{3;\pi}(x,\mu)$  and the other with a pseudo-tensor
structure $\phi^\sigma_{3;\pi}(x,\mu)$. They can be derived from light-cone
QCD sum rules (LCSR) and are usually expressed as truncated conformal series 
expansion over Gagenbauer polynomials \cite{BraunFi,CHPT1,Ball}. Their 
asymptotic forms are given by
\begin{eqnarray}
\label{eq:DAs-asy}
\phi^{({\rm as})}_{2;\pi}(x)=\frac{3f_{\pi}}{\sqrt{2N_c}}\,x(1-x)\quad;\quad\phi^{{p}\,({\rm as})}_{3;\pi}(x)=\frac{f_{\pi}}{2\sqrt{2N_c}}\quad;\quad\phi^{{\sigma}\,({\rm as})}_{3;\pi}(x)=\frac{3f_{\pi}}{\sqrt{2N_c}}\,x(1-x)\,,
\end{eqnarray}
where $f_\pi\approx 131$ MeV is the pion decay constant
and $x$ is the longitudinal momentum fraction of the valence partons.
The intrinsic TMD of the total pion wavefunctions is modeled via the
Brodsky-Huang-Lepage (BHL) prescription \cite{BHL} having the general
impact ``$b$''-space representation:
\begin{eqnarray}
\tilde{{\mathcal P}}_{t;\pi}(x,b,\mu,{\mathcal
  M}_q)\!=\!A_{t;\pi}\,\phi_{t;\pi}(x,\mu)\,\,{\rm{exp}}\left[-\frac{\beta^2_{t;\pi}{\mathcal
      M}^2_q}{x(1-x)}\right]\,{\rm{exp}}\left[-\frac{b^2x(1-x)}{4\beta^2_{t;\pi}}\right]\,;\,t=2,3\,\,,
\end{eqnarray}
where $\phi_{t;\pi}(x,\mu)$ is one of the twist-2 or twist-3 non-asymptotic 
collinear DAs at any given scale $\mu$. Such DAs satisfy
the Efremov-Radyushkin-Brodsky-Lepage (ER-BL) evolution equation
\cite{CollFactor}, e.g., the twist-2 DA is given at the LO by the
following non-asymptotic expression, in terms of Gagenbauer 
polynomials $C^{3/2}_n(2x-1)$:
\begin{eqnarray}
\phi_{2;\pi}(x,\mu)= \phi^{\rm as}_{2;\pi}(x)\sum^{\infty}_{n=0,2,4,\cdots} a^{\pi}_n(\mu^2_0)\,\,
  C^{3/2}_n(2x-1)\left(\frac{\alpha_s(\mu^2)}{\alpha_s(\mu^2_0)}\right)^{-4\gamma^{(0)}_n/9}
 + {\mathcal O}(\alpha_s)\,,
\end{eqnarray}
where $\alpha_s$ is the standard (two loop) $\overline{\rm{MS}}$ QCD coupling
with $\Lambda_{\rm QCD}\!=\!0.2$ GeV, $a^{\pi}_n$'s are the moments of the 
DA, depicting the genuine non-perturbative inputs, and $\gamma^{(0)}_n$'s 
are the corresponding standard LO anomalous dimensions. A compilation of the 
numerical values of the Gagenbauer moments as well as the LO RGE behavior of the
various non-perturbative parameters of the twist-2 and twist-3 DAs can 
be found in \cite{Raha,Ball}, normalized to the mass scale $\mu_0=1$ GeV.

The BHL Gaussian parameters $A_{t;\pi}$ and $\beta_{t;\pi}$ are fixed using 
phenomenological constraints from $\pi^0\rightarrow\gamma\gamma$ and  
$\pi\rightarrow \mu \nu_\mu$ decays (see, e.g., \cite{Raha}),
and ${\mathcal M}_q\approx 0.33$ GeV is the constituent ($q=u,d$)
quark mass, introduced to parameterize
the QCD vacuum effects. These parameters could be additionally
constrained using a combined analysis of data from lattice simulations and
from experiments like  CLEO, BaBar and FermiLab E791 diffractive dijet
production. However, since the analysis \cite{Raha} showed that the
sensitivity to the  model DA parameters is less than $5\%$, while the 
experimental error bars are much larger, we refrain from doing 
such a involved analysis at the moment. With the availability of higher 
quality data in future such a systematic combined analysis may provide 
important constraints to our results.

Next, using the TMD modified factorization ansatz in the operator convolution form
$\{F,G\}^{\rm hard}_\pi\sim \tilde{{\mathcal P}}_{\pi}\otimes{\mathcal M}_{\rm
  LO}\otimes\tilde{{\mathcal P}}_{\pi}$, where the LO matrix
elements ${\mathcal M}_{\rm LO}$ are diagrammatically represented in the
Fig.\ref{fig:feynman} and $\otimes$ represents the phase space
integration, one can obtain the pQCD contribution to the
hard form factor in a standard way up to twist-3 corrections, given by
$\{F,G\}^{\rm hard}_\pi(Q^2)=\delta \{F,G\}^{({\rm twist}2)}_\pi(Q^2)+
\delta \{F,G\}^{({\rm twist}3)}_\pi(Q^2)$ where,
\begin{eqnarray}
\label{eq:ff-hard2}
\delta \{F,G\}^{({\rm twist}2)}_\pi(Q^2)\!&=&\!\frac{64\pi}{3}\, Q^2\,\int^1_0
dxdy\int^{\infty}_{0} b_1db_1 b_2 db_2\,\alpha_s(t)\,\left[\pm x\,{\mathcal P}_{2;\pi}(x,b_1)\,{\mathcal P}_{2;\pi}(y,b_2)\right]\nonumber\\
&&\times\,{\mathcal{H}_\pm}(x,y,Q,b_1,b_2)\,S_t(x)\,{\rm exp}\left[-S(x,y,b_1,b_2,Q)\right]\,;
\end{eqnarray}
\begin{eqnarray}
\label{eq:ff-hard3}
\delta \{F,G\}^{({\rm twist}3)}_\pi(Q^2)\!&=&\!\frac{128\pi}{3}\,\mu^2_\pi\,\int^1_0
dxdy\int^{\infty}_{0} b_1db_1 b_2db_2\,\alpha_s(t)\nonumber\\
&&\hspace{-3.6cm}\times\,\left[{\bar x}\,{\mathcal
    P}^{\,p}_{3;\pi}(x,b_1)\,{\mathcal
    P}^{\,p}_{3;\pi}(y,b_2)+\frac{(1+x)}{6}\,\partial_{x}{\mathcal
    {P}}^{\,\sigma}_{3;\pi}(x,b_1)\,{\mathcal P}^{\,p}_{3;\pi}(y,b_2)+\frac{1}{2}{\mathcal P}^{\,\sigma}_{3;\pi}(x,b_1)\,{\mathcal P}^{\,p}_{3;\pi}(y,b_2)\right]\nonumber\\
&&\times\,{\mathcal H}_\pm(x,y,Q,b_1,b_2)\,S_t(x)\,{\rm exp}\left[-S(x,y,b_1,b_2,Q)\right]\,.
\end{eqnarray}
In the above equations, ``$+$'' and ``$-$'' correspond to the space-like and
time-like cases, respectively, 
${\mathcal P}_{t;\pi}(x,b)\equiv\tilde{\mathcal P}_{t;\pi}(x,b,1/b,{\mathcal{M}}_{u,d})$
and $t\!=\!{\rm max}(\sqrt{x}\,Q,1/b_1,1/b_2)$ is related to the factorization 
scale. The so-called ``chiral'' parameter $\mu_\pi$ arises from the standard 
definitions of the twist-3 collinear DAs defined at a suitable low energy scale,
$\mu_\pi(\mu_0\approx 1\,{\rm GeV})\!=\!m^2_\pi/(m_u+m_d)\!\sim\!1.7$ GeV 
\cite{BraunFi}. However, in the context of intermediate energies, $\mu_\pi$ 
is usually taken to be slightly lower $\approx 1.3-1.5$ GeV which is 
consistent with fits to the $B\rightarrow\pi$ transition form factors 
\cite{KLS,Sanda,Li,B-decay}, $\chi$PT estimates \cite{CHPT1,CHPT2} and 
the moment calculation applying QCDSR \cite{HWZ}. Since, the twist-3 results
can be somewhat sensitive to this parameter, here we use $\mu_\pi=1.5\pm
0.2$ GeV and indeed show that its variation contributes
to a large uncertainty in the time-like region. The hard kernels 
${\mathcal H}_\pm$ could be expressed in terms of the standard Bessel 
functions $K_0(\theta), \,I_0(\theta),\, 
H^{(1)}_0(\theta)=J_0(\theta)+iY_0(\theta)$ and $J_0(\theta)$:
\begin{eqnarray}
\label{eq:kernel_s}
{\mathcal H}_{+}(x,y,Q,b_1,b_2)\!&=&\!K_0(\sqrt{xy}\,Qb_2)\nonumber\\
&&\times\left[\theta(b_1-b_2)K_0(\sqrt{x}\,Qb_1)I_0(\sqrt{x}\,Qb_2)\right.\nonumber\\
&&+\left. \theta(b_2-b_1)K_0(\sqrt{x}\,Qb_2)I_0(\sqrt{x}\,Qb_1)\right]\,;
\end{eqnarray}
\begin{eqnarray}
\label{eq:kernel_t}
{\mathcal H}_{-}(x,y,Q,b_1,b_2)\!\!&=&\!\!\left(\frac{i\pi}{2}\right)^2H^{(1)}_0(\sqrt{xy}\,Qb_2)\nonumber\\
&&\times\left[\theta(b_1-b_2)H^{(1)}_0(\sqrt{x}\,Qb_1)J_0(\sqrt{x}\,Qb_2)\right.\nonumber\\
&&+\left. \theta(b_2-b_1)H^{(1)}_0(\sqrt{x}\,Qb_2)J_0(\sqrt{x}\,Qb_1)\right]\,,
\end{eqnarray}
The Sudakov factor $S(Q)$ and the jet function $S_t(x)$ are introduced to
organize to all orders the large double logarithms $\alpha_s{\rm ln}^2k_{T}$ 
($k_{T}$ is the generic transverse parton momenta) and $\alpha_s{\rm ln}^2x$, 
respectively, that arise from radiative gluon effects and may otherwise 
invalidate perturbative factorization. Such resummations result in the natural suppression of 
possible non-perturbative and kinematic endpoint enhancements of the scattering
kernel, thereby, improving convergence and making perturbative evaluation self-consistent. 
For their explicit expressions, one is referred to \cite{St,LiSt,Stefanis,Sanda,Li,Raha}. 
A few comments regarding our factorized results (Eqs.~\ref{eq:ff-hard2} 
and \ref{eq:ff-hard3}) are now in order:

1) Here, we have presented a LO analysis of the hard kernel which is
apparently gauge dependent (light-cone gauge), arising from the contribution
of the single gluon propagator. However, in \cite{NandiLi}
it was shown that for the $\pi\gamma^*\rightarrow\gamma$ transition form
factor, the gauge invariance of the hard kernel is a consequence of the
gauge-dependence cancellation between the quark level diagrams of the full QCD and
effective diagrams of the pion wavefunction, order by order in perturbation
theory using the principle of mathematical induction. In this way, the hard
kernel and the resulting predictions from the $k_T$-factorization turn out
to be gauge-invariant to all orders. The above reference also claims
that such an approach could be extended to other elastic and transition form
factors, at least up to the level of NLO corrections. 

2) Our result for the hard form factor depends on the renormalization/factorization
scale which is typical of all fixed order calculations. The Sudakov factor 
that resums a certain class of radiative soft-gluon contributions to all orders in
perturbation theory is inherently factorization scale dependent, while the 
LO hard kernel that is used to evaluate the hard form factor depends on the
renormalization scale through the running of $\alpha_s$. In this case, the 
scale dependence is minimized by adhering to a fixed prescription with the 
renormalization/factorization scale set to the momentum transfer $Q$ 
\cite{NandiLi,Coriano}. It is, however, believed that a systematic higher 
order calculation can eventually absorb this scale dependence.

3) There may be a simple rationale why the TMD factorization is expected
to work at the level of $1/Q^2$ power suppressed corrections, although a more 
rigorous proof is beyond the scope of this paper. Firstly, note that the
``active'' soft gluons which may arise e.g., from the 3-particle twist-3 DA
that probe the hard kernel, bring about additional power corrections. Compared
to the 2-particle twist-3 corrections considered in this work, the 3-particle 
twist-3 corrections is not chirally enhanced (there is a large parametric enhancement 
from $\mu_\pi$ in the definition of the 2-particle twist-3 DAs, 
which brings about a sensitivity to the chiral scale), and should be numerically
small. Secondly, the rest of the ``long-distance'' soft gluons that do not interfere 
with the hard kernel may break the TMD factorization. However, in the
large $Q^2$ limit, a hadron tends to have a small ``color-dipole'' due to the
Lorentz contraction and the Sudakov suppression. Such gluons can not
probe the small ``color-dipole'' configurations of $q\bar{q}$ within the
hadronic bound state, and their effects cancel each other. This is the
so-called ``color transperancy hypothesis''. With this assumption, one 
only needs to care about collinear gluon effects and their factorization. 
Using similar arguments, the authors in \cite{LiNagashima} have
explicitly proven TMD factorization at the twist-2 level and collinear 
factorization at the twist-3 level. Hence, it is our assumption that the 
approach presented in the above reference can even be straightforwardly 
extended to include the twist-3 TMD factorization.

\begin{figure}[t]
    \hspace{0cm}
      \resizebox{13.5cm}{!}{\includegraphics{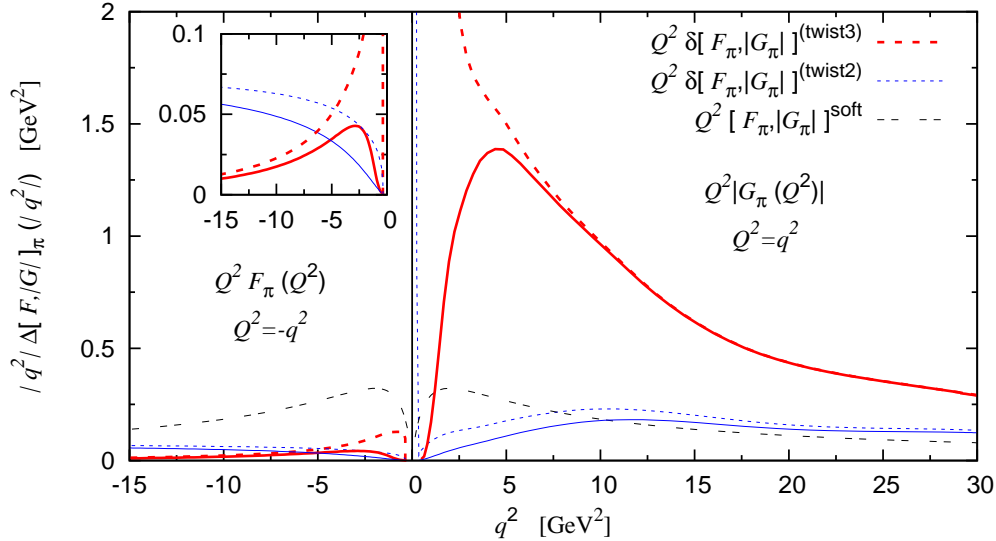}}
    \caption{
      The relative magnitudes of the soft (double-dot black lines), twist-2
      (thin solid blue lines) and twist-3 (thick solid red lines) corrections to
      the pion form factor. The twist-2 and twist-3 corrections without
      including the pre-factors are also displayed.}
    \label{fig:Dpi_ff}
\end{figure}

\begin{figure}[t]
    \hspace{0cm}
      \resizebox{13.5cm}{!}{\includegraphics{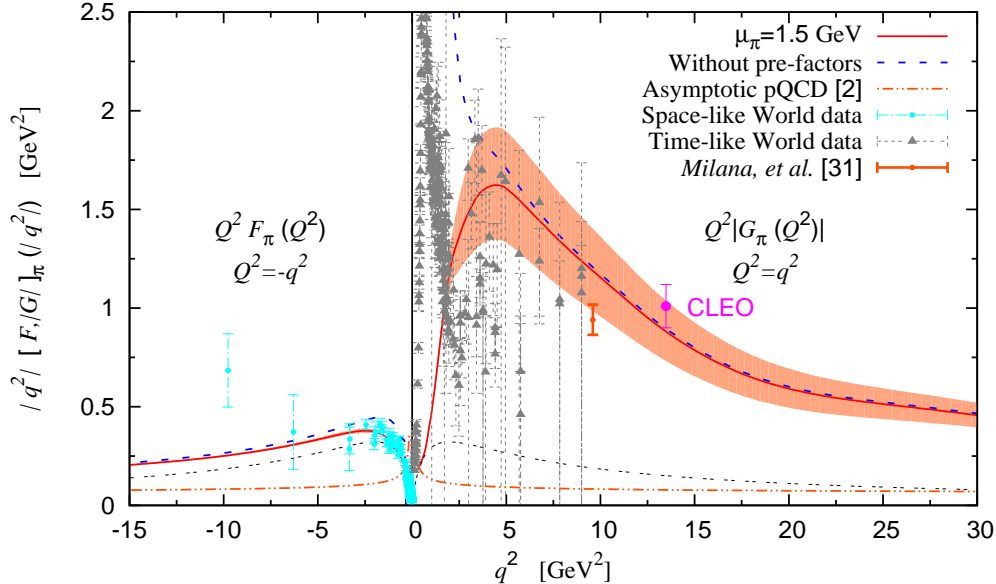}}
    \caption{
      The total space- and time-like e.m.\,\,pion form factors calculated using
      Eq.~\ref{eq:total_ff}, denoted by the solid (red) lines; the soft from factor
      $\{F,|G|\}^{\rm soft}_{\pi}$ is denoted by the double dashed (black) lines; 
      the Ward-identity violating result that does not include the pre-factor 
      modification is also displayed. For comparison, the standard asymptotic pQCD
      result \cite{BF} is displayed. The shaded area is roughly our estimated 
      theoretical error beyond $Q^2\approx5$ GeV$^2$. The world pion data are 
      taken from \cite{ee-hh1,ee-hh2,CLEO,pion_exp}. }
    \label{fig:pi_ff}
\end{figure}

\noindent {\it Soft QCD.\,} Next, following \cite{Bakulev}, we include the soft
(Feynman mechanism) contribution via Local Duality (LD) for the space-like
form factor \cite{Radyushkin},
\begin{equation}
\label{eq:ff-sp-soft}
F^{\rm soft}_\pi(Q^2)|_{\rm LD}=1-\frac{1+6s_0/Q^2}{(1+4s_0/Q^2)^{3/2}}\,,
\end{equation}
where $s_0\approx 0.68$ GeV$^2$ is the duality interval for higher excited
and continuum thresholds which is very naturally almost the ``middle'' between
pion mass $m^2_\pi\approx0$ and that of the $A_1$ resonance
$m^2_{A_1}\approx1.6$ GeV$^2$. The VMD models and $\chi$PT predictions
are not expected to work beyond $\approx 1$ GeV, while standard pQCD
with only twist-2 operators completely fails to explain the available
experimental data. The soft contribution, on the other hand, is
significantly large at moderate energies \cite{Bakulev} and so are the twist-3
power corrections \cite{Raha}. However, both the soft and the twist-3
corrections are expected to fall off rapidly as $\sim 1/Q^4$ for large $Q$,
so that asymptotically ($Q\rightarrow\infty$) one recovers the rigorous
leading twist-2 contributions $\sim 1/Q^2$ which dominate the form factor.
This aspects is clearly revealed through our analysis
(see, Figs.~\ref{fig:Dpi_ff} and \ref{fig:pi_ff}). 

To extend the analysis to the time-like region, one may analytically continue
Eq.~\ref{eq:ff-sp-soft} from the space-like region. Using such a model ansatz,
the authors of \cite{BSRadyushkin} were able to show for the first time
a much larger contribution to the form factor in the time-like region than in the
space-like, and hence were partly able to resolve the bulk of the discrepancy
for large $Q$. However, this gives rise to a single pole at
$4s_0\!\approx\!2.71$ GeV$^2$  which does not correspond to any
of the real physical bound states or resonances (e.g, $\rho,\omega$,...) 
seen in the time-like data. In fact, the observed spectrum around 2.7 GeV 
already appears to be rather ``smooth'' and well above the resonance region 
(below $\pi\omega$ threshold). Hence, rather than trying to reproduce the 
actual time-like data, including the various bound states and resonances, 
we try to explain the continuum  contribution with a smooth $G^{\rm soft}_\pi$ 
which has the same leading $1/Q^{4}$ dependence as $F_{\pi }^{\mathrm{soft}}$ 
under analytic continuation. Thus, we choose the same form of the time-like 
soft factor, i.e., $G^{\rm soft}_\pi(Q^2)=F^{\rm
  soft}_\pi(Q^2)+\mathcal{O}(1/Q^6)$ for large $Q$. Here again we should stress 
that our analysis is entirely based on the assumption that the physically 
observed low-lying resonances would not spoil the continuum contribution which 
appear as ``superposed peaks'' on a continuum spectrum.

\noindent {\it Results and Discussion.} To this end, it may be notable that the
rather {\it ad hoc} incorporation of the soft part from QCDSR have 
no {\it a priori} correspondence with the hard parts, and therefore, may lead to 
the possibility of some double-counting between the respective soft 
and hard contributions at the intermediate regime. However, such double
counting could partly be removed by imposing the the vector Ward-identity 
$\{F,G\}_\pi(0)=1$. Following the argument detailed in \cite{Bakulev,Raha}, 
we introduce appropriate power correcting pre-factors to restore the
Ward-identity, and hence we arrive at our final expression for the space- 
and time-like pion form factors given by
\begin{eqnarray}
\label{eq:total_ff}
\{F,G\}_\pi(Q^2)=1-\frac{1+6s_0/Q^2}{(1+4s_0/Q^2)^{3/2}}&+&\Delta \{F,G\}^{({\rm twist}2)}_\pi(Q^2)+\Delta \{F,G\}^{({\rm twist}3)}_\pi(Q^2)\,;\nonumber\\
\Delta \{F,G\}^{({\rm
    twist}2)}_\pi(Q^2)\!&=&\!\left(\frac{Q^2}{2s_0+Q^2}\right)^2 \delta
\{F,G\}^{({\rm twist}2)}_\pi(Q^2)\,,\nonumber\\
\Delta \{F,G\}^{({\rm
    twist}3)}_\pi(Q^2)\!&=&\!\left(\frac{Q^4}{4s^2_0+Q^4}\right)^2 \delta \{F,G\}^{({\rm twist}3)}_\pi(Q^2)\,.
\end{eqnarray}

The above pre-factors of $\delta\{F,G\}^{({\rm twist}2)}_\pi$ 
and $\delta\{F,G\}^{({\rm twist}3)}_\pi$
ensure a ``smooth'' matching of the different power-law $Q^2$
behavior between the soft and the hard parts that preserve the gauge
invariance condition $\{F,G\}^{\rm hard}_\pi(0)=0$. In principle,
this vector Ward identity can also be achieved with larger $n$ values
in the $Q^{2n}/\left( \left(2s_{0}\right) ^{n}+Q^{2n}\right)$ factors in
front of the hard part. However, as $n\rightarrow \infty $, the factor
becomes a step function which is not smooth. Thus, we have chosen the
minimum $n$'s to achieve the maximum smoothness. The individual contributions
of the soft $\{F,|G|\}^{\rm soft}_{\pi}$, twist-2 $\Delta \{F,|G|\}^{\rm (twist2)}_{\pi}$ and 
twist-3 $\Delta \{F,|G|\}^{\rm (twist3)}_{\pi}$ are summarized in Fig.~\ref{fig:Dpi_ff}.
The soft and the twist-3 terms turn out to give dominant contributions
at the low and moderate range of $Q^2$-values with anomalously large twist-3
contributions in the time-like region. Nevertheless, both the corrections exhibit
sharp fall-off with increasing $Q^2$, such that the plot extended beyond  
$Q^2\approx50$ GeV$^2$, will clearly show the twist-2 contributions as the 
being the only dominant ones, both for the space- and time-like domains. Note that
our factorized hard results are calculated using non-asymptotic collinear 
twist-2 and twist-3 DAs (up to NLO in conformal twist), taken from {\it Ball et al.}
\cite{Ball} where these are obtained in the context of LCSR.  

The final result for the total space- and time-like form factors
(Eq.~\ref{eq:total_ff}) is displayed in Fig.~\ref{fig:pi_ff}, along with the 
existing experimental pion data. The solid (red) curves correspond to our
central result obtained with the chiral parameter $\mu_\pi=1.5$ GeV, while the 
shaded area can be regarded as our estimated theoretical error, reliable only 
beyond the resonance region. Due to the overwhelming time-like twist-3 
contributions at intermediate energies, the modulus of the total time-like
form factor $\left\vert G_{\pi }\right\vert$ shows a big enhancement, at least 
by a factor of $\approx 2-4$ compared to $F_{\pi }$, although both clearly
show asymptotic trends, numerically approaching the standard pQCD result 
\cite{BF} (beyond $\sim 50-100$ GeV$^2$). The above enhancement is largely 
due to two model independent features:

(1) The quark and gluon propagators in the hard kernel can become
on-shell at non-zero $q^{2}$ in the time-like but not the
space-like region. Thus, in general
$\left\vert G_{\pi}^{\mathrm{hard}}\right\vert$ should be bigger than
$F_{\pi }^{\mathrm{hard}}$. In pQCD, this generic feature is captured
only if the $k_{T}$ dependence is kept, so that the denominator of the hard
kernel has terms proportional to $q^{2}$ and $k_{T}^{2}$ that cancel each
other in the time-like but not the space-like region
(see, e.g., Eqs.~27 and 43 of \cite{Raha} before they are Fourier
transformed into Eq.~\ref{eq:kernel_s} of this paper). Without the $k_{T}$
dependence, this enhancement will be missing.

(2) The twist-3 contributions are more important than the twist-2 ones
for intermediate range of $Q^{2}$ due to two notable reasons: Firstly,
the twist-3 terms have the aforementioned parametric enhancement arising from 
$\mu_\pi$ which is absent in the twist-2 case. Secondly, the finiteness of 
$\phi_{3;\pi }^{\,{p}}$ and the derivative of $\phi_{3;\pi}^{\,{\sigma}}$ with
respect to $x$ (see, Eq.~\ref{eq:ff-hard3}) at the endpoints $x=0,1$. 
These features together with $H\propto x^{-3/2}$,  being oscillatory 
in the time-like region and exponentially decaying in the space-like 
region (i.e., the time-like parton propagators in momentum representation
develop poles which are absent in the space-like), account for 
the characteristic relative enhancement of the time-like twist-3 contributions.

With these two rather robust features and typical treatments of the soft and
sub-leading twist-3 contributions, it appears that the previously reported
discrepancy between the experimental data and theoretical predictions
\cite{CLEO} can be ostensibly reconciled. To this end, we also present our 
results without including the pre-factors to demonstrate their effect.
As revealed from  Figs.~\ref{fig:Dpi_ff} and \ref{fig:pi_ff}, 
without the pre-factors the hard contributions tend to grow very rapidly as 
$Q^2\rightarrow0$ and become unreliable, while at the same time beyond $Q^2\approx5-10$ 
GeV$^2$ the effect of the pre-factors is hardly discernable. Clearly, then our 
predictions convincingly agrees with most of the space- and time-like 
experimental pion data, including the recent CLEO result:
$Q^{2}|G_{\pi }(13.48$ GeV$^{2})|=1.01\pm 0.11$(stat)$\pm0.07$(syst)
GeV$^2$ \cite{CLEO}, and also the theoretical 
prediction $M_{J/\psi }^{2}\left\vert G_{\pi }(M_{J/\psi
}^{2}\!=\!9.6\,\mathrm{GeV}^{2})\right\vert=0.94\pm0.08$
GeV$^2$ \cite{Milana}, fixed from branching ratios of 
$J/\psi \rightarrow \pi \pi $ and $J/\psi \rightarrow e^+ e^-$ decays.

Finally, to comment on the error estimate of our approach, we first look at the error
band in Fig.~\ref{fig:pi_ff}. While the width of the error band is too narrow
to be even noticeable in the space-like region, it is anomalously large in the
time-like region. Over $90\%$ of this error is essentially due to the
variation of the chiral parameter $\mu_\pi$ between $1.3-1.7$ GeV with
increasing contribution to the pion form factor. The remaining difference generously 
over-estimates the other model (parameter) dependences in the DAs from QCDSR, 
but it seems to be a reasonable range of theoretical error when the error of
the soft part is also included. However, we again stress that the estimate
only applies beyond the resonance region. Furthermore, several 
aspects deserve to be noted: The extent of the theoretical error from our LO 
analysis is large enough to completely subsume the systematic errors that may arise, 
e.g., considering NLO effects (in the QCD coupling $\alpha_s$)
\cite{Stefanis,NLO1,NLO2}, sub-leading twists (see, e.g., \cite{twist-46}
for the twist-4 and twist-6 contributions to the 
pion form factor in the context of QCDSR), and effects due to higher
Fock state corrections which are expected to be rather nominal. For 
example, even without explicit calculations, it is easily understandable 
that the 2-particle twist-4 power corrections are, in fact, very small 
being being proportional to $m^2_\pi\rightarrow 0$. Again, the contribution of
the 3-particle twist-3 DA, being proportional to the ``tiny'' non-perturbative 
parameter $f_{3\pi}\approx 0.45\times 10^{-2}$ GeV$^2$ (to be compared with
the 2-particle twist-3 DA parameter $\mu_\pi\approx1.5$ GeV), is also strongly 
suppressed. Thus, the 2-particle twist-3 contributions are indeed very special
in this regard. Moreover, it is estimated that the NLO corrections in the case of the 
$\pi\gamma^*\rightarrow\gamma$ transition form factor amount to only about $5\%$ 
under specific factorization scheme with the factorization scale set to the 
energy/momentum transfer $Q$ \cite{NandiLi}. This is not expected to be very
different for the pion form factor. To conclude, the unnaturally 
large twist-3 contribution, especially in the time-like region, is certainly 
non-intuitive and may constitute an important step toward understanding the 
large asymmetry seen in the experimental data, unaccountable otherwise.  

The authors thank C.-W Kao, J. Qiu, A. V. Radyushkin,
W. Schroers, N. G. Stefanis, Z. T. Wei and X.-G. Wu for fruitful
discussions. We are especially thankful to H.-N. Li for various constructive 
comments. JWC thanks the KITPC, Beijing for hospitality.



\end{document}